\documentclass[12pt]{article}

\usepackage{hyperref}

\usepackage{graphics,psfrag}
\usepackage{amsthm,amssymb,epsfig,amsmath,euscript,array,cite}
\usepackage{color}
\usepackage{graphicx}
\usepackage{amscd}
\usepackage{slashed}

\newcommand{\be}{\begin{equation}}
\newcommand{\ee}{\end{equation}}
\def\bea{\begin{eqnarray}}
\def\eea{\end{eqnarray}}

\newcommand{\eq}[1]{\begin{equation}#1\end{equation}}
\newcommand{\spl}[1]{\begin{split}#1\end{split}}

\def\nn{\nonumber}

\newcommand{\beq}{\begin{equation}}
\newcommand{\eeq}{\end{equation}}
\newcommand{\ben}{\begin{eqnarray}}
\newcommand{\een}{\end{eqnarray}}
\newcommand{\bes}{\begin{subequations}}
\newcommand{\ees}{\end{subequations}}
\newcommand{\blg}{\begin{align}}
\newcommand{\elg}{\end{align}}

\newcommand{\ao}{\mathring{A}}
\newcommand{\co}{\mathring{C}}
\newcommand{\go}{\mathring{g}}

\def\one{\mbox{1 \kern-.59em {\rm l}}}
\def\dphi1{{\dot\phi_1}}
\def\dphi2{{\dot\phi_2}}
\def\dphi3{{\dot\phi_3}}
\def\dphi{{\dot\phi}}

\def\={\, =\, }
\def\d{\delta}    
\def\e{\epsilon}

\makeatletter \@addtoreset{equation}{section} \makeatother

\begin{document}

\begin{titlepage}
\hfill MCTP-16-34
\vspace{14pt}

\begin{center}

{\Large \bf Supersymmetric  IIB Background  with $AdS_4$ Vacua }\\

\vspace{.4cm}

{\Large \bf from Massive  IIA Supergravity}\\

\vspace{1.6cm}

{\bf \footnotesize Leopoldo A. Pando Zayas$^{a,}$\footnote{lpandoz@umich.edu, $^2$tsimpis@ipnl.in2p3.fr, $^3$catherine.whiting@wits.ac.za}, Dimitrios Tsimpis${^{b,2}}$, and Catherine A. Whiting$^{c,3}$ }

\vspace{1cm}

{\bf }

\vspace{.2cm}

\vspace{.4cm}
{\it ${}^a$  Michigan Center for Theoretical Physics,  Department of Physics}\\
{\it University of Michigan, Ann Arbor, MI 48109, USA}\\

\vspace{.4cm}

{\it ${}^b$ Universit\'e Claude Bernard (Lyon 1)}\\
{\it UMR 5822, CNRS/IN2P3, Institut de Physique Nucl\'eaire de Lyon  }\\
{\it 4 rue Enrico Fermi, F-69622 Villeurbanne Cedex, France  }\\

\vspace{.4cm}

{\it ${}^c$ National Institute for Theoretical Physics}\\
{\it School of Physics and Mandelstam Institute for Theoretical Physics}\\
{\it University of the Witwatersrand, Johannesburg}\\
{\it WITS 2050, South Africa}

\vspace{14pt}

\end{center}
\begin{abstract}
We present a new Type IIB supergravity background of the warped form AdS$_4\times \mathcal{M}_6$  with dilaton, $B$-field and all Ramond-Ramond fluxes turned on.  We obtain the solution by  applying non-Abelian T-duality to a certain representative of a class of AdS$_4$ backgrounds in massive  IIA supergravity. By explicitly constructing the Killing spinor of the seed solution  and using an argument involving Kosmann spinorial Lie derivative we demonstrate that the background is supersymmetric. 
\end{abstract}

\end{titlepage}


\section{Introduction}

One of the most distintive characteristics of string theory in its aim to connect with the real world is the need for compactification. The search for four-dimensional vacua has, consequently, a distinguished and rich history. Although originally the focus has been on compactifications on Calabi-Yau spaces leading to supersymmetric Minkowski vacua, in the last decade the tools for understanding compactification with fluxes have been developed; for a review see \cite{Grana:2005jc}.  In this direction $AdS_4$ vacua play a central role as they are considered a potentially important stepping stone toward de Sitter vacua. One of our motivations in this work is to widen the class of $AdS_4$ vacua not just by presenting one new background but by demonstrating the reach of a particular solution-generating mechanism. Another motivation for the study of $AdS_4$ vacua arises from holography where such supergravity solutions are conjectured to be dual to three-dimensional conformal field theories. 

The main new ingredient that we exploit to construct the new solution is non-Abelian T-duality (NATD)\cite{Ossa:1992vc,Fridling:1983ha,Fradkin:1984ai} . There has recently been a resurgence of interest in non-Abelian T-duality   including its systematic extension to the Ramond-Ramond sector \cite{Sfetsos:2010uq, Lozano:2011kb}. One immediate application of this duality has been  to generate solutions from various seed backgrounds in the context of the AdS/CFT correspondence, for a limited list see \cite{Itsios:2013wd, Lozano:2012au,Lozano:2014ata, Bea:2015fja, Lozano:2015bra, Lozano:2015cra}.

One of the hopes that we have in pursuing the construction of explicit solutions using non-Abelian T-duality is that it could provide new solutions that have avoided classification efforts in constructing $AdS$ vacua. One such example is a  new solution with $AdS_5$ factor \cite{Macpherson:2014eza} that defied previous classification schemes \cite{Gauntlett:2005ww} which assumed non-vanishing $F_5$ flux. This  solution stimulated  work to go back and complete  the classification efforts providing further insight in some new classes of solutions \cite{Couzens:2016iot}. This last effort brought about understanding into a class of $AdS_5$ solutions with no D3 brane interpretation. This is an example of the synergy between NATD and the general structure of $AdS$ vacua in supergravity. We hope that our efforts in this manuscript  might lead to similar scrutiny in the important series of classifying solutions with an $AdS_4$ factor, see for example \cite{Lust:2009zb}.

Several  new solutions with $AdS_4$ factors were recently provided in \cite{Zayas:2015azn}; in this manuscript we extend to a more interesting class by applying non-Abelian T-duality to a representative background  in the massive type IIA class presented in \cite{Lust:2009mb}. 
These are  $\mathcal{N}=2$ solutions of the form  AdS$_4\times \mathcal{M}_6$, where the internal manifold $\mathcal{M}_6$ is 
locally a codimension-one foliation such that the five-dimensional leaves admit a Sasaki-Einstein structure. 
Alternatively, $\mathcal{M}_6$ can be thought of as a two-sphere bundle over a four-dimensional K\"{a}hler-Einstein base.

The rest of the manuscript is organized as follows. In the next section \ref{Sec:Review} we review the class of massive IIA solutions of \cite{Lust:2009mb}; we also discuss its ``massless''  limit which provides a perhaps more intuitive form of the solution. In this section we also present explicitly the Killing spinors which will be subsequently used to argue for the supersymmetry of the dual solution. In section \ref{Sec:NATD} we present the non-Abelian T-dual solutions corresponding to the massless limit and to the full massive solution in Type IIB. We conclude in section \ref{Sec:Conclusions}.


\section{Massive $AdS_4$ backgrounds}\label{Sec:Review}

Let us start by reviewing the solutions corresponding to the class described in \cite{Lust:2009mb}.  
These are massive type IIA $\mathcal{N}=2$ solutions of the form  AdS$_4\times \mathcal{M}_6$, with $\mathcal{M}_6$ 
a two-sphere bundle $S^2(\mathcal{B}_4)$ over a four-dimensional K\"{a}hler-Einstein base $\mathcal{B}_4$.  
In the case where $\mathcal{B}_4$ is a smooth manifold of 
positive curvature,\footnote{Smooth four-dimensional K\"{a}hler-Einstein manifolds of 
positive curvature were classified in \cite{Tian:1987if}: they are $\mathbb{CP}^1\times\mathbb{CP}^1$, $\mathbb{CP}^2$, 
and the del Pezzo surfaces $dP_3,\dots,dP_8$.} these solutions can be thought of  as massive IIA deformations 
of the $\mathcal{N}=2$ IIA circle reductions of 
the M-theory  AdS$_4\times Y^{p,q}(\mathcal{B}_4)$ backgrounds of \cite{Gauntlett:2004hh,Martelli:2008rt}, where $Y^{p,q}(\mathcal{B}_4)$ is a seven-dimensional Sasaki-Einstein manifold. 
The first such massive deformation was constructed in \cite{Petrini:2009ur} and corresponds to the special case $p=2$, $q=3$, $\mathcal{B}_4=\mathbb{CP}^2$ 
 (the $Y^{3,2}(\mathbb{CP}^2)$ 
space is also referred to as $M^{1,1,1}$ in the physics literature).\footnote{A closely related AdS$_4\times S^6$ massive IIA solution is that of \cite{Guarino:2015jca}. Although $S^6$ can locally be put in the form of an $S^2$ bundle over $\mathbb{CP}^2$, its topology is different from 
that of any regular $S^2(\mathcal{B}_4)$ bundle. This can be seen e.g. by comparing their second Betti numbers. Other AdS$_4$ backgrounds in massive IIA have been discussed in \cite{Varela:2015uca}.}

We will adopt the conventions of \cite{Tsimpis:2012tu} which 
result in certain simplifications. 
The general form of this solution is: 
\bea
\label{MM}
ds^2_{10}& =& e^{2A(\theta)}ds^2(AdS_4)+ds^2(\mathcal{M}_6),\nn \\
ds^2(\mathcal{M}_6)&=&e^{2C(\theta)}ds^2(\mathcal{B}_4)+e^{2A(\theta)}(f^2(\theta)d\theta^2+\sin^2\theta(d\Psi+\mathcal{A})^2)~,
\eea
with
\be
f(\theta)=\frac{1}{2-\sin^2\theta e^{2(A-C)}}~,
\ee
where the two warp factors $A$, $C$ obey the following system of first-order 
differential equations:
\eq{\spl{\label{7}
A'&=\frac{1}{2}\tan\theta\frac{1-\sin^2\!\theta ~\!e^{2(A-C)}}{2-\sin^2\!\theta ~\!e^{2(A-C)}},\\
C'&=\frac{1}{4}\sin(2\theta)\frac{
e^{2(A-C)}}{2-\sin^2\!\theta ~\!e^{2(A-C)}}
\frac{1+e^{8A}}{1+\cos^2\theta e^{8A}}
~,}}
where a prime denotes a derivative with respect to $\theta$. For convenience, we drop the explicit $\theta$ dependence in $A(\theta)$ and $C(\theta)$.

We now specialize to the case where the base manifold is $\mathcal{B}_4=\mathbb{CP}^2$. In this particular case the metric is given by 
\be
\label{B_4}
ds^2(\mathcal{B}_4)=3 \big(d\mu^2+\frac{1}{4}\sin^2\mu(\sigma_1^2+\sigma_2^2+\cos^2\mu \sigma_3^2)\big)~,
\ee
where we use the following definitions for SU(2) Maurer-Cartan forms,
\bea \label{sigmai}
\sigma_1&=&\sin\psi d\theta_2 -\cos\psi\sin\theta_2 d\phi_2, \nonumber \\
\sigma_2&=&\cos\psi d\theta_2 +\sin\psi \sin\theta_2 d\phi_2, \nonumber \\
\sigma_3&=& d\psi +\cos\theta_2 d\phi_2.
\eea
The connection that appears in Eq. (\ref{MM})  is  $\mathcal{A}=-\frac{3}{4}\sin^2\mu\,\,\sigma_3$ and it is related to the K\"{a}hler form $j$ on $\mathcal{B}_4$ by $d\mathcal{A}=-j$.  The metric on 
$\mathcal{B}_4$ is normalized so that $R_{mn}=2g_{mn}$.

The NS flux is written in terms of the RR fluxes, to ensure the Bianchi identities $dF_p-H\wedge F_{p-2}=0$ are obeyed, which leads to
\be\label{1}
B_2=\beta+\frac{1}{F_0}F_2,
\ee
where $\beta$ is some closed 2-form. The dilaton is given by,
\be
e^{2\phi}=\frac{e^{6A}}{1+\cos^2\theta e^{8A}}~.
\ee
The RR fluxes are given by,\footnote{The RR fluxes, given here in the conventions of the democratic formalism, have all legs along the internal space: they are the so-called `magnetic' 
fluxes. There are also `electric' fluxes which are related to the above by ten-dimensional Hodge duality.}
\bea
F_0&=&-1,\nn \\
F_2&=& \frac{e^{2C-4A}}{\cos\theta}Vj-e^{-2A}Z_1(d\Psi+\mathcal{A})\wedge d\theta, \nn \\
F_4&=& \frac{e^{4C}}{2}\hat{V}j\wedge j+2e^{2A+2C}\cos\theta Z_2j\wedge (d\Psi+\mathcal{A})\wedge d\theta,\nn \\
F_6&=& -\frac{3e^{4C-2A}}{2}\sin\theta f(\theta)j\wedge j\wedge (d\Psi+\mathcal{A})\wedge d\theta,
\eea
where we have defined $V=(1-\sin^2\theta e^{2(A-C)})$, $\hat{V}=(2V-1)$, $Z_1=(\sin\theta+2A'\cos\theta)$, and $Z_2=(\cos\theta A'-\sin\theta)$ and taken $g_s=1,\ L=1$ for simplicity.

\subsection{Vanishing Romans mass limit}
\label{MasslessLimitSec}

In the following it will be useful to take the limit of zero Romans mass, which 
we will call the ``massless'' limit of the solution. This solution has been presented in \cite{Petrini:2009ur}.
In order to do so we must first reinstate $g_s$ and $L$ in the solution. 
Moreover we define:
\eq{\label{ACDef}\mathring{g}_s:=L^{-3}g_s~;~~~e^{\mathring{A}}:=L e^A~;~~~e^{\mathring{C}}:=L e^C~.}
With these definitions the metric reads:
\bea\label{mt}
ds^2_{10}& =& e^{2\ao}\widehat{ds}^2(AdS_4)+ds^2(\mathcal{M}_6),\nn \\
ds^2(\mathcal{M}_6)&=&e^{2\co}ds^2(\mathcal{B}_4)+e^{2\ao}
(f^2(\theta)d\theta^2+\sin^2\theta(d\Psi+\mathcal{A})^2)
\eea
with
\be
f(\theta)=\frac{1}{2-\sin^2\theta e^{2(\ao-\co)}}~,
\ee
where now $\widehat{ds}^2(AdS_4)$ is the metric of an AdS space with {\it unit} 
radius of curvature. 
The two warp factors $A$, $C$ obey the following system of first-order 
differential equations:
\eq{\spl{\label{8}
\ao'&=\frac{1}{2}\tan\theta\frac{1-\sin^2\!\theta ~\!e^{2(\ao-\co)}}{2-\sin^2\!\theta ~\!e^{2(\ao-\co)}}\\
\co'&=\frac{1}{4}\sin(2\theta)\frac{
e^{2(\ao-\co)}}{2-\sin^2\!\theta ~\!e^{2(\ao-\co)}}
\frac{1+L^{-8}e^{8\ao}}{1+L^{-8}\cos^2\theta e^{8\ao}}
~.}}
The NS flux is written in terms of the RR fluxes as before,
\be
B_2=\beta+\frac{1}{F_0}F_2,
\ee
where $\beta$ is some closed 2-form.  The dilaton is given by,
\be\label{dt}
e^{2\phi}=\frac{\go_s^2e^{6\ao}}{1+L^{-8}\cos^2\theta e^{8\ao}}.
\ee
The RR fluxes  are given by,
\bea\label{rrflux}
\go_sF_0&=&-L^{-4},\nn \\
\go_sF_2&=& \frac{e^{2\co-4\ao}}{\cos\theta}Vj-e^{-2\ao}Z_1(d\Psi+\mathcal{A})\wedge d\theta, \nn \\
\go_sF_4&=& L^{-4}\frac{e^{4\co}}{2}\hat{V}j\wedge j+2L^{-4}e^{2\ao+2\co}\cos\theta Z_2j\wedge (d\Psi+\mathcal{A})\wedge d\theta,\nn \\
\go_sF_6&=& -\frac{3e^{4\co-2\ao}}{2}\sin\theta f(\theta)j\wedge j\wedge (d\Psi+\mathcal{A})\wedge d\theta,
\eea
where we have again defined $V=(1-\sin^2\theta e^{2(\ao-\co)})$, $\hat{V}=(2V-1)$, $Z_1=(\sin\theta+2\ao'\cos\theta)$, and $Z_2=(\cos\theta \ao'-\sin\theta)$.


The massless limit consists in taking,
\eq{L\rightarrow\infty~;~~~\go, \ao, \co\rightarrow\mathrm{finite}~.}
Taking this limit in Eq. (\ref{8}) we see that the warp factors obey the system,
\eq{\spl{\label{9}
\ao'&=\frac{1}{2}\tan\theta\frac{1-\sin^2\!\theta ~\!e^{2(\ao-\co)}}{2-\sin^2\!\theta ~\!e^{2(\ao-\co)}},\\
\co'&=\frac{1}{4}\sin(2\theta)\frac{
e^{2(\ao-\co)}}{2-\sin^2\!\theta ~\!e^{2(\ao-\co)}}
~.}}
It is worth pointing out that the system in Eq. (\ref{8}) could also be studied perturbatively around the 
point $L=\infty$, along the lines of \cite{Petrini:2009ur}. This is akin to the perturbation provided for explicit examples in \cite{Gaiotto:2009yz}.  The meaning of turning on a Romans mass was established to be dual to choosing, in the context of ABJM, a non-vanishing sum of Chern-Simons levels  \cite{Gaiotto:2009mv}.  It would be, therefore, quite interesting to study the system in Eq. (\ref{8}) more generally.  

 The system in Eq. (\ref{9}) admits the following analytic solution
\eq{\label{sol}\ao=\co=-\frac{1}{4}\log(2(1+\cos^2\theta))+A_0~,}
where $A_0$ is a constant, which also gives,
\eq{f(\theta)=\frac{1}{1+\cos^2\theta}~.}
The metric is as in Eq. (\ref{mt}), which is manifestly independent\footnote{Note that under the coordinate transformation,
\eq{\cos\theta=\frac{\cos t}{\sqrt{1+\sin^2t}}~,}
the metric takes the form,
\eq{
ds^2_{10}= e^{2A_0}\sqrt{\frac{1+\sin^2t}{2}}\Big[\widehat{ds}^2(AdS_4)
+ds^2(\mathcal{B}_4)
+\frac{1}{2}d t^2+\frac{2\sin^2t}{1+\sin^2t}(d\Psi+\mathcal{A})^2\Big]
~,} which, after a rescaling of the $\Psi$ coordinate, matches the metric presented in \cite{Petrini:2009ur, Zayas:2015azn}.
}  of $L$,
\eq{\label{MasslessMetric}
ds^2_{10}= \frac{e^{2A_0}}{\sqrt{2}\sqrt{1+\cos^2\theta}}\Big[\widehat{ds}^2(AdS_4)
+ds^2(\mathcal{B}_4)
+\frac{1}{(1+\cos^2\theta)^2}d\theta^2+\sin^2\theta(d\Psi+\mathcal{A})^2\Big]
~.}
The RR fluxes $F_2$ and $F_6$ are also manifestly independent of $L$, as given in Eq. (\ref{rrflux}). They can be written explicitly in a manifestly closed form,
\bea\label{rrfluxm}
\go_sF_2&=&-e^{-2A_0}d\big[
{\sqrt{2}(1+\cos^2\theta)^{\frac12}\cos\theta}(d\Psi+\mathcal{A})
\big]\nn \\
\go_sF_6&=& -\frac{3}{2}e^{2A_0}
d\big[
{\frac{1}{\sqrt{2}}(1+\cos^2\theta)^{-\frac12}\cos\theta}
j\wedge j\wedge (d\Psi+\mathcal{A})\big]~.
\eea
Moreover we have $F_0=F_4=0$ in the massless limit, and also $H=0$.\footnote{Note that Eq.~(\ref{1}) is no longer valid in the massless limit: For nonzero Romans mass  the $F_2$ Bianchi identity can be used to solve for $H$, leading to Eq.~(\ref{1}). On the other hand in the case of zero Romans mass the $F_2$ Bianchi identity reads $dF_2=0$ and does not impose any constraints on $H$. In other words Eq.~(\ref{1}) should simply be discarded in the massless limit, and be replaced by the condition that $F_2$ is closed.} Finally, the dilaton is given by,
\eq{\label{MasslessDilaton}
e^{2\phi}=\go_s^2e^{6\ao}
~,}
as follows from Eq. (\ref{dt}).
This background can be thought of as arising from a dimensional reduction of $AdS_4\times M^{1,1,1}$ along the $\phi_1$ angle and was studied by Petrini and Zaffaroni in \cite{Petrini:2009ur}. The non-Abelian T-duality of this background was presented in \cite{Zayas:2015azn}.

\subsection{The Killing spinors and supersymmetry}
We now proceed to explicitly present the supersymmetry of the background of Eq. (\ref{MM}) and its massless limit.
The $SU(3)$ structure in the massless limit is constructed as follows. First we 
define a complex one-form $K$ and a local $SU(2)$ structure $(\tilde{j}, \tilde{\omega})$,
\eq{\spl{\label{ls}
K&=e^{\ao}\left[\frac{id \theta}{1+\cos^2\theta}-\sin\theta(d\Psi+\mathcal{A}) \right],\\
\tilde{j}&=e^{2\ao}\left(\sin\theta~\!\Re\omega+\cos\theta~\! j\right),\\
\tilde{\omega}&=e^{2\ao}\left(\cos\theta~\!\Re\omega-\sin\theta~\! j+ i\Im\omega\right)~,
}}
where $\ao$ is given in Eq. (\ref{sol}), $j$ is the K\"{a}hler form 
 of $\mathcal{B}_4$,  $\omega:=e^{2i\Psi}\hat{\omega}$, and $\hat{\omega}$ is the holomorphic 
two-form of $\mathcal{B}_4$ which satisfies,
\eq{j\wedge\hat{\omega}=0~;~~~j\wedge j=\Re\hat{\omega}\wedge\Re\hat{\omega}=\Im\hat{\omega}\wedge\Im\hat{\omega}~;~~~
d\hat{\omega}=2i\mathcal{A}\wedge\hat{\omega}~.}
Then one can construct an $SU(3)$ structure $(J,\Omega)$ on $\mathcal{M}_6$ given by,
\eq{\label{JOmega}\Omega=i\tilde{\omega}\wedge K~;~~~J=\tilde{j}+\frac{ i}{2} K\wedge K^*~.}
As noted in \cite{Lust:2009mb} there is  a one-parameter family of $SU(3)$ structures on $\mathcal{M}_6$, 
which are obtained from Eqs.~(\ref{ls}), (\ref{JOmega}) by $\theta$-independent $SO(2)$ rotations in the
$(\Re\omega,\Im\omega)$ plane.
%
%
These rotations 
act nontrivially on the almost complex structure of
$\mathcal{M}_6$ (and thus on the Killing spinor, as we explain below) while leaving invariant the
metric and the fluxes of the solution. 
The upshot is that the background possesses a 
one-parameter family of Killing spinors associated with the $SO(2)$  family of $SU(3)$ structures generated by Eq. (\ref{JOmega}), 
consistent with the $\mathcal{N} = 2$ supersymmetry of the solution.

The Killing spinor  is obtained as follows. 
We first note that associated to an $SU(3)$ structure on $\mathcal{M}_6$ there is a (generally non-integrable) almost 
complex structure which  can be constructed out of $\Re\Omega$ alone \cite{Hitchin:2000jd}, 
\eq{\label{hit}I_m{}^n=\frac{1}{24} \varepsilon^{np_1\dots p_5}~\!\Re\Omega_{mp_1p_2}\Re\Omega_{p_3p_4p_5}~,}
where $\varepsilon^{p_1\dots p_6}$ is purely numeric and the normalization constant was fixed by imposing $I_m{}^pI_p{}^n=-\d_m^n$. On the 
other hand there is a correspondence between almost complex structures and Weyl spinors, up to complex multiplication.\footnote{More precisely,  
on a $2n$-dimensional Riemannian spin manifold $\mathcal{M}_{2n}$ there is a correspondence between almost complex structures and line bundles of pure spinors. However, 
for $n\leq3$ every Weyl spinor is pure.} 
The (positive-chirality, internal part of the) Killing spinor of the supergravity solution is precisely the Weyl spinor associated to the almost complex structure Eq. (\ref{hit}) which is, in its turn, induced by the $SU(3)$ structure.  

The way to explicitly construct the spinor associated to an almost complex structure is described in some detail in e.g.~\cite{Tsimpis:2016bbq}. Here we will outline the main steps. 
Let us define a holomorphic/antiholomorphic projector with respect to the non-integrable almost complex structure,
\eq{\big[\Pi^{\pm}\big]_m{}^n:=\frac12(\delta_m^n\mp i I_m{}^n)~.}
Correspondingly, we define holomorphic/antiholomorphic gamma matrices as follows,
\eq{\gamma^{\pm}_m:=\big[\Pi^{\pm}\big]_m{}^n\gamma_n ~,}
with $\gamma_m:=e_m{}^a\gamma_a$, where $\gamma_a$ are 
gamma matrices in flat six-dimensional space\footnote{In constructing the explicit Killing spinor solutions, we will use the 6d gamma matrix basis, \bea
&\gamma_1=\sigma_0\otimes \sigma_2\otimes\sigma_1, \quad 
\gamma_2=\sigma_0\otimes \sigma_2\otimes\sigma_3, \quad 
\gamma_3=\sigma_1\otimes \sigma_0\otimes\sigma_2,\nn\\&
\gamma_4=\sigma_3\otimes \sigma_0\otimes\sigma_2, \quad 
\gamma_5=\sigma_2\otimes \sigma_1\otimes\sigma_0, \quad 
\gamma_6=\sigma_2\otimes \sigma_3\otimes\sigma_0.\nn
\eea}.  
From the definition above it follows that,
\eq{
\{ \gamma^{\pm}_m, \gamma^{\pm}_n\}=0~;~~~\{ \gamma^{\pm}_m, \gamma^{\mp}_n\}=2\big[\Pi^{\pm}\big]_{mn}=g_{mn}\mp i J_{mn}
~.}
The Killing spinor $\eta$ is then given as a solution to the algebraic equation,
\eq{\label{Killing}\gamma^-_m\eta=0~.}

Starting from the  $SU(3)$ structure in Eq.~(\ref{JOmega}) and following the procedure described above,  
a family of solutions to Eq. (\ref{Killing}), parameterized by one real parameter $p$,  can be shown to be the following:
\be
\eta=\frac{1}{\alpha}
\left(
\begin{array}{c}
  1+\sin \theta  \sin 2 \Psi +p (\sin \theta  \cos 2 \Psi + i\cos\theta) \\
 p(-1+ \sin \theta  \sin 2 \Psi )  -\sin \theta \cos 2 \Psi +i\cos\theta \\
 -p\cos\theta +i (1+ p \sin \theta  \cos 2 \Psi + \sin \theta  \sin 2 \Psi ) \\
 -\cos\theta +i(p(-1+\sin \theta  \sin 2 \Psi )+\sin \theta  \cos 2 \Psi ) \\
 p(-1+ \sin \theta  \sin 2 \Psi) -\sin \theta  \cos 2 \Psi +i\cos\theta) \\
 -1- \sin \theta \sin 2 \Psi -p (\sin \theta  \cos 2 \Psi +i\cos\theta) \\
 -\cos\theta+i (p (-1 +\sin \theta  \sin 2 \Psi )-\sin \theta  \cos 2 \Psi ) \\
 p\cos\theta-i( 1+\sin \theta \sin 2 \Psi +p\sin \theta  \cos 2 \Psi ) \\
\end{array}
\right)~.
\ee
The value of the normalization $\alpha$ is given by:
\be
\alpha=2 \sqrt{2}  \sqrt{1+p^2+2p\sin \theta \cos 2\Psi-(p^2-1)\sin\theta\sin 2\Psi},\nn
\ee
and it ensures that $\eta^{\dagger}\eta=1$.  
Of course the solution to Eq. (\ref{Killing}) can only be determined up to complex multiplication: it gives rise to 
a one-parameter family of spinor line bundles which should correspond to the $SO(2)$ family of SU(3) structures generated from Eq. (\ref{JOmega}) in the way described above.  
The (internal part of the) full Killing spinor is given by $\Theta=c e^{A/2}\eta$, where $c$ is a complex constant and $A$ is the warp factor.
Clearly, the Killing spinor $\Theta$ depends only on the spacetime coordinates $\theta$ and $\Psi$ which are coordinates of the $S^2$ fiber.

It is intuitively clear that if we dualize along directions on which the Killing spinor does not depend, we will preserve supersymmetry. In our case the Killing spinor is  independent of the coordinates ($\theta_2,\phi_2,\psi$) defining the SU(2) isometry on which we dualize; we can conclude that supersymmetry is preserved after applying NATD.  This is precisely the claim made in \cite{Zayas:2015azn} to argue for the supsersymmetry of the background in Eq. (\ref{NATD(MasslessLimit)}).  A way to make the intuitive argument regarding independence of coordinates rigorous was first presented in \cite{Sfetsos:2010uq}. Clearly, what is required is to turn the coordinate-dependent statement into a tensorial, coordinate-independent statement. Indeed, \cite{Kelekci:2014ima} established that the supersymmetry variations after T-duality are related to the variations before T-duality through the Kosmann spinorial Lie derivative, which vanishes when the Killing spinors are independent of the T-duality directions, thus providing a rigorous basis for the intuition alluded to in our reasoning.

For the Killing spinors in the massive case we follow a very similar procedure to the one  above, but with the following modification to the local SU(2) structure forms,
\eq{\spl{
K&=e^{A}\left[i f(\theta) d\theta -\sin\theta(d\Psi+\mathcal{A}) \right],\\
\tilde{j}&=e^{2C}\left(\sin\theta~\!\Re\omega+\cos\theta~\! j\right),\\
\tilde{\omega}&=e^{2C}\left(\cos\theta~\!\Re\omega-\sin\theta~\! j+ i\Im\omega\right)~,
}}
where $\omega:=e^{i\zeta(\theta)}e^{2i\Psi}\hat{\omega}$ with $f$, $\zeta$ functions of $\theta$ which we do not need to specify explicitly. 
This case possesses dynamic SU(3) structure, therefore there will be two independent spinors $\eta_1$, $\eta_2$ on $\mathcal{M}_6$ corresponding to the two $SU(3)$ structures constructed from $\Omega_1=i\omega\wedge K,\ \Omega_2=-i\omega^*\wedge K$.  
The (internal parts of the) full Killing spinors are then certain $\theta$-dependent linear combinations of $\eta_1$, $\eta_2$. 
In the following it will be convenient to define $\tau:=\zeta (\theta )+2 \Psi$.

An explicit  solution to the massive Killing spinor Eq. (\ref{Killing}) is given by,
\be \label{Killing1}
\eta_1=\frac{1}{\alpha_1}
\left(
\begin{array}{c}
 1+\sin \theta  \sin \tau+p \sin \theta  \cos \tau+i p \cos \theta  \\
-p -\sin \theta \cos \tau+p \sin \theta  \sin \tau+i \cos \theta \\
 -p\cos \theta +i (1+ \sin \theta  \sin \tau+ p \sin\theta  \cos\tau)\\
 -\cos \theta +i(-p- \sin \theta  \cos\tau+ p \sin \theta  \sin \tau) \\
-p -\sin \theta \cos\tau+ p\sin \theta \sin\tau+i \cos \theta \\
 -1-\sin \theta \sin\tau-p \sin \theta \cos \tau-i p \cos \theta  \\
-\cos \theta+i(-p - \sin \theta  \cos \tau + p \sin \theta  \sin \tau)\\
 p\cos \theta-i( 1+\sin \theta  \sin \tau+ p \sin\theta  \cos \tau )
   \\
\end{array}
\right).
\ee
The normalization $\alpha_1$ is given by 
\be
\alpha_1=2\sqrt{2} \sqrt{1+p^2+2 p\sin \theta  \cos \tau-\left(p^2-1\right)\sin\theta\sin\tau}\nn.
\ee
The second, linearly independent,  Killing spinor solution is,
\be \label{Killing2}
\eta_2=\frac{1}{\alpha_2}
\left(
\begin{array}{c}
-1 +\sin \theta  \sin\tau+p \sin \theta  \cos \tau+i p \cos \theta  \\
p -\sin \theta  \cos\tau+p \sin \theta  \sin \tau+i \cos \theta  \\
-p \cos \theta+ i (-1+\sin \theta  \sin\tau+ p \sin\theta  \cos \tau ) \\
-\cos \theta+i(p - \sin \theta  \cos \tau +p \sin \theta  \sin \tau)  \\
 p-\sin \theta \cos \tau+p\sin \theta \sin \tau +i \cos \theta \\
1 -\sin \theta  \sin \tau-p \sin \theta \cos \tau-i p \cos \theta  \\
-\cos \theta +i(p- \sin \theta  \cos \tau +p \sin \theta  \sin \tau) \\
p \cos \theta +i(1 - \sin \theta  \sin \tau-p \sin\theta  \cos \tau)\\
\end{array}
\right),
\ee
with normalization 
\be
\alpha_2=2 \sqrt{2} \sqrt{1+p^2-2 p \sin \theta  \cos\tau+\left(p^2-1\right) \sin \theta  \sin \tau}.\nn
\ee
As it is clear form the explicit expressions Eq.'s (\ref{Killing1}) and (\ref{Killing2}) the Killing spinors in the massive case are also independent of the coordinates along which the NATD is performed and thus supersymmetry is preserved in the massive dual solution, which we will present in section \ref{NATD(Massive)}. 

\section{Type IIB Backgrounds from Non-Abelian T-duality }\label{Sec:NATD}
The non-Abelian T-duality procedure has been utilized to generate many new supergravity backgrounds.  Its application to the background in Eq.'s (\ref{mt}) - (\ref{rrflux}) constitutes one of our main results.  Hence, a brief review of the procedure (specific to the dualization along SU(2) isometries) \cite{Itsios:2013wd} is included for the benefit of the reader.  
\\
Essentially, the non-Abelian T-duality procedure is a generalization of what is done in Abelian T-duality.  A 3-step B{\"u}scher procedure is applied to a 2d string $\sigma$ model whose target space possesses a non-Abelian isometry group.  Throughout the paper we will assume SU(2) so that the NS sector fields can be written as,
\bea  
\label{generic metric}
ds^2&=&G_{\mu\nu}(x)dx^{\mu}dx^{\nu}+2G_{\mu i}(x)dx^{\mu}L^i+g_{ij}(x)L^iL^j,\nn \\
B&=&B_{\mu\nu}(x)dx^{\mu}\wedge dx^{\nu}+B_{\mu i}(x)dx^{\mu}\wedge L^i+\frac{1}{2}b_{ij}(x)L^i\wedge L^j
\eea
where $\mu,\nu = 1,...7$  and $i,j=1,2,3$.  Here the $L^i$'s are Maurer-Cartan forms, which are writen explicitly in Eq. (\ref{sigmai}) above as $\sigma_i$.  
We will write the Lagrangian density succinctly as,
\begin{equation}
\label{L0short}
\mathcal{L}_0=Q_{AB}\partial_{+} X^A\partial_{-}X^B,
\end{equation}
where $A,B=1,...,10$ and
\begin{equation}
Q_{AB}=
\left(
\begin{array}{c|c}
\\ \ Q_{\mu\nu}\quad & Q_{\mu i} \\ \\  \hline \
 Q_{i\mu} \quad  & E_{ij} \\
\end{array}
\right),
\quad  \partial_{\pm}X^A=\left(\partial_{\pm}X^{\mu},\ L_{\pm}^i\right),
\end{equation}
with \begin{equation}
Q_{\mu\nu}=G_{\mu\nu}+B_{\mu\nu},\quad Q_{\mu i}=G_{\mu i}+B_{\mu i},\quad Q_{i\mu}=G_{i \mu}+B_{i\mu},\quad E_{ij}=g_{ij}+b_{ij}.
\end{equation}
The first step is to gauge the SU(2) isometry by changing derivatives to covariant derivatives and introduce gauge fields.  Next, one needs to ensure the gauge fields are non-dynamical and can be integrated out, therefore, the second step is to add a Lagrange multiplier term to Eq. (\ref{L0short}).  Three new variables (corresponding to the dimension of SU(2)) are introduced in the form of Lagrange multipliers, $v_i$.  Eliminating the angles of SU(2) and adopting the three Lagrange multipliers as new coordinates is a convenient gauge fixing choice.
Step three is to integrate out the gauge fields to obtain the dual Lagrangian density,
\begin{equation}
\hat{\mathcal{L}}=\hat{Q}_{AB}\partial_{+} \hat{X}^A\partial_{-}\hat{X}^B \label{dualL},
\end{equation}
where we can read off the dual components of $\hat{Q}_{AB}$ from,
\begin{equation}
\hat{Q}_{AB}=
\left(
\begin{array}{c|c}
\\ \ Q_{\mu\nu}-Q_{\mu i}M_{ij}^{-1}Q_{j\nu}\quad & Q_{\mu j}M_{ji}^{-1} \\ \\  \hline \
 -M_{ij}^{-1}Q_{j\mu} \quad  & M_{ij}^{-1} \\
\end{array}
\right),
\quad  \partial_{\pm}\hat{X}^A=\left(\partial_{\pm}X^{\mu},\ \partial_{\pm}v^i\right).
\end{equation}
We have additionally defined
$M_{ij}=E_{ij}+f_{ij}$, and $f_{ij}=\alpha'\e_{ij}^{\ \ k}v_k$.
The dual metric and $\hat{B}_2$ are the symmetric and antisymmetric components of $\hat{Q}_{AB}$, respectively. The dilaton transformation is given by
\begin{equation}
\label{dualdilaton}
\hat{\Phi}=\Phi-\frac{1}{2}\text{ln}(\frac{\text{det}M}{\alpha'^3}).
\end{equation}

The transformation of the RR fluxes is formally quite elegant, though in practice can be cumbersome.  First, a bispinor is constructed out of the RR forms and their Hodge duals, (in Type IIA):
\be
P=\frac{e^{\Phi}}{2}\sum^5_{n=0}\slashed{F}_{2n},\quad \slashed{F}_p=\frac{1}{p!}\Gamma_{\mu_1...\mu_p}F_p^{\ \mu_1...\mu_p}
\ee
Next, we construct a matrix $\Omega$ defined by,
\be \Omega=\frac{(\Gamma^1\Gamma^2\Gamma^3+\zeta_a\Gamma^a)\Gamma_{11}}{\sqrt{\alpha'^3}\sqrt{1+\zeta^2}},\ee
with $\zeta^a=\kappa^a_{\ i}z^i$, where $\kappa$ and $z$ are determined by the original geometry and Lagrange multipliers by  $\kappa^a_{\ i}\kappa^a_{\ j}=g_{ij}$ and  $z^i=\frac{1}{\det \kappa}(b^i+v^i)$.
Finally, the dual fluxes simply arise from inverting $\Omega$:
\be
\hat{P}=P \Omega^{-1}.
\ee
\subsection{NATD of Massless Limit}
\label{NATD(MasslessLimit)}
In this section we present the background resulting from applying non-Abelian T-duality along the $SU(2)$ isometry defined by the $\sigma_i$ in Eq. (\ref{sigmai}) on the background given by Eq.'s (\ref{MasslessMetric}) - (\ref{MasslessDilaton}).  This background was originally presented in \cite{Zayas:2015azn}, but we present it here to match our notation and normalization conventions for convenience.  
As argued in section 2.1 above, the background is manifestly independent of the scale $L$.  We will continue to use the notation introduced in that section, noting that the explicit form of $\ao$ is given by Eq. (\ref{sol}) above.  The $\alpha'$ terms are introduced by the NATD via the Lagrange multipliers, $v_i$. We have chosen NATD gauge fixing $v_i\to m x_i$, transformed to spherical coordinates, ($\rho,\chi,\xi$), and conveniently fixed $m=\frac{3}{8}$. 

The NS sector of this background is
\bea \label{MasslessNS}
\hat{ds}^2&=& e^{2\ao}ds^2(AdS_4) + e^{2\ao}(3 d\mu^2+\frac{1}{(1+\cos^2\theta)^2}d\theta^2)\nn \\&~&+\frac{4e^{2\ao}\sin^2\theta\cos^2\mu}{Q}d\Psi^2+\frac{3}{4M}e^{2\ao}\alpha'^2\sin^2\mu d(\rho\sin\chi)^2 \nn \\&~&+\frac{81}{4096\Delta}\bigg[ \frac{ e^{4\ao}\rho^2\sin^4\mu\sin^2\chi}{ \alpha' Q}d\xi\Psi^2\nn \\&~&+\bigg(\frac{\alpha'^{3/2}\rho^2\cos\chi\sin\chi }{ \sqrt{M} }d(\rho\sin\chi)- \sqrt{\frac{M}{\alpha' }} d(\rho\cos\chi)\bigg)^2\bigg],  \nn \\
\hat{B}_2&=&\frac{81 e^{2 \ao} \rho^2 \sin ^2\mu \sin\chi}{8192 Q \Delta}d\xi\Psi\wedge d\rho\chi +\frac{3\alpha'\sin^2\theta}{2Q}d(\rho\cos\chi)\wedge d\Psi,\nn \\
e^{-2\hat{\Phi}}&=& \frac{\Delta}{\go_s^2e^{6\ao}},\quad \Delta=\frac{27e^{2\ao}\sin^2\mu}{1024\alpha'^3}\big[4e^{4\ao} \sin^4\mu Q +  \alpha'^2 \rho^2  K\big],
\eea
where we have defined the following one-forms,
\bea
d\xi\Psi&=& \big(Q d\xi-4\sin^2\theta d\Psi\big),\nn \\
d\rho\chi&=&\Big(\rho  K d\chi +\cos\chi \sin\chi(Q-4) d\rho\Big),\nn \\
d\theta\mu&=&\big(Z_1\sin\mu d\theta-2\cos\mu\cos\theta d\mu\big),
\eea
and included the following definitions,
\bea
Q&=&4\cos^2\mu+3\sin^2\mu\sin^2\theta,\nn \\
K&=&Q \cos^2\chi+4\sin^2\chi,\nn \\
M&=&\alpha'^2\rho^2\cos^2\chi+4e^{4\ao}\sin^4\mu.
\eea
The RR sector contains all of the fluxes, given by
\bea \label{MasslessRR}
\go_s\hat{F}_1&=&\frac{9e^{-2\ao}\sin\mu}{32 \sqrt{\alpha'}}\bigg[\cos\theta\sin\mu d(\rho\cos\chi)  -\rho\cos\chi d\theta\mu\bigg],\nn \\
\go_s\hat{F}_3&=&\frac{9e^{-2\ao}\sqrt{\alpha'}\rho \cos\mu }{32Q}\bigg[3\cos\theta\sin^2\theta\sin\mu d\mu+2\cos\mu Z_1 d\theta\bigg] \wedge d\rho\wedge d\Psi \nn \\&~& 
+\frac{729\rho^3\sin^3\mu\sin\chi}{262144\sqrt{\alpha'}Q\Delta} \bigg[\cos\theta\cos\chi\sin\mu d\chi-4\sin\chi d\theta\mu \bigg]\wedge d\rho\wedge d\xi\Psi\nn \\&~&
-\frac{27e^{2\ao}\cos\mu\sin^3\mu\sin\theta}{4a(\theta)\alpha'^{3/2}}d\theta\wedge d\mu \wedge d\Psi \nn \\&~&
-\frac{729 e^{4\ao}\rho\sin^7\mu \sin\chi }{65536\alpha'^{5/2}\Delta}\bigg[ d\theta\mu\wedge d(\rho\sin\chi)\wedge d\xi\Psi \bigg],
\nn \\
\go_s\hat{F}_5&=&\frac{27\sqrt{\alpha'}\rho}{64 }d\text{Vol}(AdS_4)\wedge d\rho \nn \\&~&+\frac{9e^{4\ao}\sin^3\mu}{16 \alpha'^{3/2}\sin\theta a(\theta)}d\text{Vol}(AdS_4)\wedge\Big(2\cos\theta\sin^2\theta\sin\mu d\theta- a(\theta)^2\cos\mu Z_1 d\mu  \Big) 
\nn \\&~&-\frac{729 e^{4\ao}\rho^2\cos\mu\sin^5\mu\sin\chi}{65536 \alpha'^{3/2}a(\theta)\Delta}\bigg[6\sin\theta d\theta\wedge d\mu\wedge d\rho\chi \nn \\&~&+a(\theta)\sin\mu\big(3\cos\theta\sin^2\theta \sin\mu d\mu+2\cos\mu Z_1 d\theta\big)\wedge d\rho\wedge d\chi \bigg]\wedge d\xi\wedge d\Psi,
\eea
where $a(\theta)=2(1+\cos^2\theta)$.

As noted in \cite{Zayas:2015azn}, this background has singularities generated by the NATD at $\mu=0$ and at $\mu=\frac{\pi}{2}$ simultaneous with $\rho=0$ or $\chi=0$.  Analogous to what happens in Abelian T-duality, the singularity at $\mu=0$ is produced by the collapsing cycle in front of the SU(2) isometry direction before the duality, as can be seen from Eq. (\ref{B_4}). The other singular locus  $\{\mu=\pi/2, \rho=0\}$ or $\{\mu=\pi/2, \chi=0\}$ is certainly milder.  

Finally, we examine the behaviour of the metric and fields near the $\mu=0$ singularity.  In this limit we find for the NS sector,
\bea \label{MasslessNS5}
ds^2&\sim& e^{2\ao}\bigg[ds^2(AdS_4)+\frac{d\theta^2}{(1+\cos^2\theta)^2}+\sin^2\theta d\Psi^2 +\frac{3}{4\nu}\Big(d\nu^2+\frac{e^{-4\ao}}{2}d\rho^2\nn \\&~&+\nu^2\big[\sin^2\chi (d\xi-\sin^2\theta d\Psi)^2+\frac{1}{\rho^2\cos^2\chi}d(\rho\sin\chi)^2\big]\Big)\bigg],\nn\\
e^{2\Phi}&\sim& \frac{256e^{4\ao}}{27 \nu \rho^2},\quad 
B_2\sim \frac{3}{8}\bigg[\sin^2\theta \cos\chi d\rho\wedge d\Psi-\rho\sin\chi d\chi\wedge d\xi\bigg],
\eea
where we have defined $\nu=\mu^2$ and set $\alpha'$ and $g_s$ to 1.  Eq. (\ref{MasslessNS5})  is consistent with the general form of NS5-branes\footnote{Type II NS5 branes are described by, \bea ds^2&=& -dt^2+dx_1^2+...+dx_5^2+H(dx_6^2+...+dx_9^2)\nn \\
e^{2\phi}&=& H,\ \ H=H(x_6,...,x_9),\ \ \nabla^2H=0\nn \eea where H is a harmonic function of the coordinates transverse to the branes.}, up to a factor of $\rho^2$ in the dilaton.  As explained in \cite{Lozano:2016wrs}, this factor arises from the differing volumes of the original and NATD submanifolds.  Thus, we determine that the $\mu= 0$ singularity is due to the presence of smeared NS5 branes, common to NATD-generated backgrounds.

For completeness, we additionally present the RR Fluxes near $\mu\sim 0$, which simplify to,
\bea
F_1&\sim & \frac{9}{32}e^{-2\ao}\bigg[\nu \Big(\cos\theta d(\rho\cos\chi)-\rho Z_1 d\theta\Big)+\rho\cos\theta\cos\chi d\nu \bigg], \nn \\
F_3 &\sim & \frac{9e^{-2\ao}\rho}{124}\bigg[\bigg(\frac{3}{2}\cos\theta\sin^2\theta d\nu+2Z_1 d\theta\bigg)\nn \\&~&+\frac{3\sin\chi}{32}\bigg(\nu\Big(\cos\theta\cos\chi d\chi-4\sin\chi Z_1 d\theta\Big)+\sin\chi\cos\theta d\nu\bigg)\bigg], \nn \\
F_5 &\sim & \frac{27\rho}{64}d\text{Vol}(AdS_4)\wedge d\rho.
\eea

It is worth noting that  \cite{Gaiotto:2014lca} provided known examples where the 6d SCFTs dual to massive Type IIA NS5, D6- D8 brane constructions have been studied in full detail.  On the supergravity side these models can further be shown to contain $AdS_4$ as a subfactor of $AdS_7$. The appearance of the $AdS_4$  further signals a dual 3d CFT theory.  The analysis of this type of 3d CFT is beyond  the scope of this work, but we hope that a more fundamental description of our solution in terms of a brane box picture will very likely mimic that of \cite{Gaiotto:2014lca} with the appropriate dualities included.

\subsection{NATD of Massive Case}\label{NATD(Massive)}
In this section we present a new Type IIB supergravity background resulting from the application of a Non-Abelian T-duality with respect to the SU(2) isometry defined by the $\sigma^i$ in $\mathcal{B}_4$ above, on the background given by Eq's (\ref{mt}) - (\ref{rrflux}).  We continue using the definitions for $\ao,\ \co$ given in Eq. (\ref{ACDef}).  
Note that we have set the closed 2-form from Eq. (\ref{1}), $\beta=0$ for simplicity.

The NS sector of the background is given by,
\bea \label{massiveNATDNS}
\hat{ds}^2&=&e^{2 \ao}ds^2(AdS_4) +e^{2\ao}f(\theta)^2d\theta^2+3e^{2\co}d\mu^2+ \frac{4 e^{2(\ao+\co)}\cos^2\mu\sin^2\theta}{\tilde{Q}}d\Psi^2\nn \\&~&
+\frac{3}{4\tilde{M}}e^{2\co}\alpha'^2\sin^2\mu d(\rho\sin\chi)^2 +\frac{81e^{2\co}\rho^2\sin^4\mu\sin^2\chi}{4096\alpha'\tilde{Q}\tilde{\Delta}}d\tilde{\xi\Psi}^2\nn \\&~&
+\frac{81}{4096 \cos^2\theta \tilde{\Delta}}\bigg[\sqrt{\frac{\tilde{M}}{\alpha'}}\cos\theta d(\rho\cos\chi)+e^{-4\ao}\bigg(\frac{2L^4}{\alpha'^{3/2}}\sin\mu\sqrt{\tilde{M}} d\tilde{\theta\mu} \nn \\&~& - \sqrt{\frac{\alpha'}{\tilde{M}}}\rho\sin\chi \big(e^{4\ao}\alpha'\rho\cos\theta\cos\chi-2e^{2\co}L^4\sin^2\mu V\big)d(\rho\sin\chi)\bigg)\bigg]^2,
\nn \\
\hat{B}_2&=&
\frac{81 e^{-4\ao}\rho\sin^2\mu\sin\chi}{8192\alpha'\cos\theta \tilde{Q}\tilde{\Delta}}\bigg[2e^{2\co}L^4\sin^2\mu V \tilde{Q} d(\rho\sin\chi) -e^{4\ao}\alpha'\rho\cos\theta d\tilde{\rho\chi}  \bigg]\wedge d\tilde{\xi\Psi}\nn \\&~&
+\frac{3\alpha'}{2\tilde{Q}}\Big(e^{2\ao}\sin^2\theta d(\rho\cos\chi)\wedge d\Psi+\frac{e^{-4\ao}L^4}{\cos\theta\sin\mu}d\tilde{\theta\mu}\wedge d(\rho\cos\chi)\Big)\nn \\&~&
+\frac{81e^{-8\ao}L^4\rho\sin\mu\sin\chi}{2048\alpha'^2\cos^2\theta\tilde{Q}\tilde{\Delta}}\bigg[-2e^{2\co}L^4\alpha'\sin^2\mu V \tilde{Q} d\tilde{\theta\mu}\wedge d(\rho\sin\chi) \nn \\&~&+e^{4\ao}\alpha'^2\rho\cos\theta d\tilde{\theta\mu}\wedge d\tilde{\rho\chi} \bigg] \nn \\
e^{-2\hat{\Phi}}&=& e^{-2\phi}\tilde{\Delta},\quad  \tilde{\Delta}= \frac{27\sin^2\mu}{1024 \alpha'^3}\left(4e^{4\co}\sin^4\mu\tilde{Q}+\alpha'^2\rho^2\tilde{K}+\tilde{\mathcal{B}}\right),
\eea
with $e^{-2\phi}$ defined in Eq. (\ref{dt}), and we have defined the following one-forms,
\bea
d\tilde{\xi\Psi}&=&\big(\tilde{Q}d\xi-4e^{2\ao}\sin^2\theta d\Psi\big),\nn \\
d\tilde{\rho\chi}&=&\big(\rho \tilde{K} d\chi +\cos\chi\sin\chi(\tilde{Q}-4e^{2\co})d\rho\big),\nn \\
d\tilde{\theta\mu}&=&\big(e^{2\ao}\cos\theta\sin\mu Z_1 d\theta-2e^{2\co}\cos\mu V d\mu \big),\nn \\
d\tilde{\mu\theta}&=&\big(e^{2\ao}\cos\theta\sin\mu Z_2 d\theta - e^{2\co}\cos\mu d\mu \big).
\eea
We have additionally defined
\bea
\tilde{Q}&=&4e^{2\co}\cos^2\mu+3e^{2\ao}\sin^2\theta\sin^2\mu,\nn \\
\tilde{K}&=& \tilde{Q}\cos^2\chi +4e^{2\co}\sin^2\chi,\nn \\
\tilde{\mathcal{B}}&=& \frac{1}{e^{8\ao}\cos^2\theta}\left(4e^{2\co}L^4\tilde{Q}\sin^2\mu V(e^{2\co}L^4\sin^2\mu V- e^{4\ao}\alpha'\rho\cos\theta\cos\chi)\right),\nn \\
\tilde{M}&=& \rho ^2 \alpha '^2 \cos ^2\chi +4 e^{4 \co}\sin ^4\mu +\frac{\tilde{\mathcal{B}}}{\tilde{Q}}. 
\eea
Note that in $\hat{B}_2$, we have omitted a total derivative term that naturally appeared in the dualization procedure. 

Before we present the RR sector of the background, we would like to make a few comments about the massless limit.  As mentioned above, we can easily track terms that vanish in the limit with the scale $L$.  
Since NATD inverts terms present in the isometry direction (analogous to the $R\to\frac{\alpha'}{R}$ in a simple case of Abelian T-duality), it would seem natural to take $L\to 0$, and $\go,\ao,\co\to$ finite in the massless limit. 
This is true for $\tilde{\Delta}$, as it reduces to $\Delta$.  In addition, the metric and $B_2$ of Eq. (\ref{massiveNATDNS}) reduce to that of Eq. (\ref{MasslessNS}).   However, the dilaton does not obey the limit because it retains the term defined in Eq. (\ref{dt}), which requires $L\to \infty$ in order to reduce to Eq. (\ref{MasslessDilaton}).  Since this is contradictory, we conclude the massless limit does not seem to exist after applying NATD.  This could be expected, given that NATD is a non-trivial, generally non-invertible transformation.
For similar reasons, we will not expect the RR fluxes to reduce to Eq. (\ref{MasslessRR}).  We will now turn to their presentation.



In the following we will define, 
\bea
\tilde{P}_1&=&(e^{8\ao}\cos^2\theta+L^8V^2),\nn \\
\tilde{P}_2&=&(e^{8\ao}\cos^2\theta \hat{V}+L^8V^2).
\eea
Then the RR Fluxes take the form,
\bea
\go_s\hat{F}_1&=& \frac{9\sqrt{\alpha'}}{64L^4}\rho d\rho 
 +\frac{9e^{-8\ao+2\co}\sin^3\mu}{16 L^4 \alpha'^{3/2}\cos^2\theta}\Big[L^8V d\tilde{\theta\mu} - 2e^{8\ao}\cos^2\theta \big(d\tilde{\mu\theta}+ 2e^{2\co}\cos\mu V d\mu  \big) \Big], \nn
\eea
\bea
\go_s\hat{F}_3&=& \frac{9e^{-2\ao+2\co}\sqrt{\alpha'}\rho\cos^2\mu Z_1}{16\tilde{Q}} d\theta \wedge d\rho\wedge d\Psi\nn \\&~&+\frac{243e^{-18\ao+8\co}\cos\mu\sin^9\mu\tilde{P}_1Z_1}{2048\alpha'^{9/2}\cos^4\theta\tilde{\Delta}}\big(L^8\tilde{Q}V^2+3e^{10\ao}\cos^2\theta\sin^2\theta\sin^2\mu\hat{V}\big)d\theta\wedge d\mu\wedge d\Psi \nn \\&~&+
 \frac{9e^{-6\ao+2\co}\cos\mu\sin\mu}{8L^3\alpha'^{3/2}\cos\theta \tilde{Q}}\bigg[\alpha'e^{2\co}\cos\mu\sin\mu(L^8VZ_1-2e^{8\ao}\cos^2\theta Z_2) d(\rho\cos\chi)\nn \\&~&-e^{4\ao}\cos\theta\big[ e^{4\ao}\alpha'\rho\cos\theta\cos\chi Z_2 \tilde{Q}-e^{2\co}L^4\sin^2\mu\big(2(4e^{2\co}\cos^2\mu+\tilde{Q})VZ_2\nn \\&~&-3f(\theta)\sin\theta\tilde{Q}\big)  \big]d\mu \bigg]\wedge d\theta \wedge d\Psi \nn \\&~& +\frac{729e^{-4\ao+2\co}L^4\rho^3\sin^3\mu}{262144\sqrt{\alpha'}\cos\theta\tilde{Q}\tilde{\Delta}}\bigg[8\cos\mu V\tilde{Q} (e^{2\co}\sin^2\chi d\xi+e^{2\ao}\cos^2\chi\sin^2\theta d\Psi)\wedge d\mu \nn \\&~&+\sin\mu\sin\chi\Big(4e^{2\ao}\cos\theta\sin\chi Z_1 d\theta -V\tilde{Q}\cos\chi d\chi\Big)\wedge d\tilde{\xi\Psi}    \bigg]\wedge d\rho\nn \\&~&
 -\frac{243e^{-8\ao+2\co}\rho^2\sin^3\mu}{131072L^4\alpha'^{3/2}\cos^2\theta\tilde{Q}\tilde{\Delta}}\bigg[\Big(3e^{2\ao}\cos\theta\sin^2\mu\sin\chi(2e^{8\ao}\cos^2\theta Z_2-L^8VZ_1) d\theta\wedge d\tilde{\xi\Psi}\nn \\&~&+6e^{2\co}\cos\mu\sin^2\mu\sin\chi\tilde{Q}\tilde{P}_2 d\mu\wedge d\xi\Big)\wedge d\tilde{\rho\chi}  -3 e^{2\co}\sin^3\mu\sin\chi\tilde{Q}\tilde{P}_1 d\rho\wedge d\chi\wedge d\tilde{\xi\Psi}  \nn \\&~&
 +8e^{2\ao}\cos\mu\cos\chi \tilde{Q}\Big(L^8\rho\cos\theta\tilde{K}VZ_1 d\theta-3e^{2\co}\sin^2\mu\sin^2\theta(2L^8V^2+\tilde{P}_2)d\rho\Big)\wedge d\mu\wedge d\Psi \bigg]\nn \\&~&
 +\frac{243e^{-14\ao+6\co}\cos\mu\sin^7\mu}{4096L^4\alpha'^{7/2}\cos^4\theta\tilde{\Delta}}\bigg[3e^{2\co}\sin^2\mu\sin^2\theta\big(2L^8e^{8\ao}\cos^2\theta V^3+L^{16}V^4+e^{16\ao}\cos^4\theta\hat{V}\big)\nn \\&~&d\mu\wedge d(\rho\cos\chi)\wedge d\Psi +L^8\rho\cos\chi\cos\theta V Z_1 \big(4e^{2\co}\cos^2\mu(\tilde{P}_1+2L^8V^2)\nn \\&~&+3e^{2\ao}\sin^2\mu\sin^2\theta(3L^8V^2+e^{8\ao}\cos^2\theta(4V-1))\big)d\theta\wedge d\mu\wedge d\Psi \bigg]\nn \\&~&
+\frac{243e^{-12\ao+3\co}L\rho\sin^5\mu}{65536\alpha'^{5/2}\cos^3\theta\tilde{Q} \tilde{\Delta}}\bigg[3e^{2\ao}\cos\theta\sin^3\mu\sin\chi V\tilde{Q}(2e^{8\ao}\cos^2\theta Z_2-L^8VZ_1)\nn \\&~& d\theta\wedge d\tilde{\xi\Psi} \wedge d(\rho\sin\chi) + 6e^{2\co}\cos\mu\sin^2\mu\sin\chi V\tilde{Q}^2\tilde{P}_2 d\mu\wedge d\xi\wedge d(\rho\sin\chi)\nn \\&~&-24e^{2\ao+2\co}\cos\mu\sin^2\mu\sin^2\theta V\tilde{Q}\big(\cos\chi\tilde{P}_2d(\rho\cos\chi)+2 V\tilde{P}_1 d\rho\big)\wedge d\mu\wedge d\Psi\nn \\&~& 
+8e^{2\ao}\rho\cos\mu\cos\theta \Big(3e^{8\ao+2\co}\cos^2\theta\sin^2\mu\sin^2\chi\tilde{Q}(2VZ_2+Z_1\hat{V})d\xi \nn \\&~&+\big[L^8V^2Z_1\tilde{Q}(3\cos^2\chi\tilde{Q}+4e^{2\co}\sin^2\chi)+3e^{10\ao}\cos^2\theta\sin^2\theta\sin^2\mu\big(\cos^2\chi\hat{V}Z_1\tilde{Q}\nn \\&~&-8e^{2\co}\sin^2\chi V Z_2\big) \big]d\Psi \Big) \wedge d\theta\wedge d\mu \bigg],\nn
\eea
\bea
\go_s\hat{F}_5&=&\frac{27\sqrt{\alpha'}}{64}\rho d\text{Vol}(AdS_4)\wedge d\rho +
\frac{9 e^{-4\ao}\sin\mu}{32L^4\sqrt{\alpha'}\cos\theta\sin\theta f(\theta)}d\text{Vol}(AdS_4)\wedge\nn \\&~& \bigg[-e^{2\co}(3L^8 f(\theta)\sin\theta V-2e^{8\ao}\cos^2\theta Z_2)\big(\sin\mu d(\rho\cos\chi) +2\rho\cos\mu \cos\chi d\mu \big) \nn \\&~& +e^{2\ao}\rho \cos\chi\cos\theta\sin\theta\sin\mu f(\theta)(e^{8\ao}f(\theta)\sin\theta\hat{V}+3 L^8 Z_1)d\theta  \bigg]
\nn \\&~&
+\frac{9e^{-8\ao+2\co}\sin^3\mu}{16 \alpha'^{3/2}\cos^2\theta\sin\theta f(\theta)}d\text{Vol}(AdS_4)\wedge \bigg[2e^{8\ao}\cos^2\theta Z_1 d\tilde{\mu\theta}-3L^8f(\theta)\sin\theta V d\tilde{\theta\mu}\nn \\&~&-2e^{8\ao}\cos\theta V\big(e^{2\ao}f(\theta)^2\sin\mu\sin^2\theta (V-1)d\theta+4e^{2\co}\cos\mu\cos\theta Z_2 d\mu\big)\bigg] \nn \\&~&
+ \frac{729e^{2\ao+4\co} \rho^3\cos\mu\sin^3\mu\sin\chi}{131072 L^4\sqrt{\alpha'}\tilde{\Delta}}\bigg[-8\cos\theta\sin\chi d\theta\wedge d\mu\nn \\&~&+\cos\chi\sin\mu\big[3\sin\mu\sin^2\theta\hat{V}d\mu-4\cos\theta\cos\mu Z_2 d\theta\big]\wedge d\chi\bigg]\wedge d\xi \wedge d\rho \wedge d\Psi   \nn \\&~&%
-\frac{729 e^{-2\ao+4\co}\rho^2\cos\mu\sin^5\mu\sin\chi}{65536\alpha'^{3/2}\cos\theta \tilde{\Delta}}\bigg[\bigg(12e^{2\co}\cos\theta\sin\theta f(\theta)\sin\chi d(\rho\cos\chi)\nn \\&~&+ 3\cos\theta\sin\theta\cos\chi\big(e^{2\ao}\sin^2\mu\sin\theta\hat{V}Z_1-f(\theta)\tilde{Q}\big) d(\rho\sin\chi)\nn \\&~&-8e^{2\co}\cos\theta V Z_2((1+\cos^2\mu)\cos\chi\sin\chi d\rho+\rho(\cos^2\mu\cos^2\chi+\sin^2\chi)d\chi)\bigg)\wedge d\theta\wedge d\mu \nn \\&~&+e^{2\co}\sin\mu(3\sin\mu\sin^2\theta V^2 d\mu-\cos\mu\cos\theta (Z_1+2VZ_2)d\theta)\wedge d\rho\wedge d\chi  \bigg]
\wedge d\xi\wedge d\Psi\nn \\&~&
 + \frac{729 e^{-6\ao+6\co} \rho\cos\mu \sin^7\mu \sin\chi }{32768L^4 \alpha'^{5/2}\cos\theta \tilde{\Delta} }\bigg[3L^8f(\theta)\sin\theta V \tilde{Q}\nn \\&~&+L^8\big(4e^{2\co}\cos^2\mu V-6e^{2\ao}\sin^2\mu\sin^2\theta(V-1)\big)Z_1\nn \\&~&+\big(8L^8e^{2\co}\cos^2\mu V^2-2e^{8\ao}\cos^2\theta\tilde{Q}\big)Z_2\bigg]d(\rho \sin\chi)\wedge d\theta\wedge d\mu \wedge d\xi\wedge d\Psi.
\eea
 For this background we have verified that the Bianchi identities $dF_1=0$ and $dF_5-H_3\wedge F_3=0$ are satisfied, which together provide non-trivial checks that this background is indeed a solution.  
 
 The singularity at $\mu=0$ is present here, again due to the collapsing cycle before duality.  Additional milder singular loci are defined by the zero's of $\tilde{\Delta}$ in Eq. (\ref{massiveNATDNS}), one at $\{\mu=\pi/2, \theta=0, \rho=0\}$ or $\{\mu=\pi/2, \theta=0, \chi=0\}$ being an obvious example.
The brane interpretation in the massive case is conceptually similar to that of the massless case in Eq. (\ref{MasslessNS5}). 
Near $\mu\sim0$, we find,
\bea
ds^2&\sim& e^{2\ao}\bigg[ds^2(AdS_4)+f(\theta)^2d\theta^2+\sin^2\theta d\Psi^2\bigg] +\frac{3}{4\nu}\bigg[e^{2\co}d\nu^2+\frac{\nu^2e^{2\co}}{\rho^2\cos^2\chi} d(\rho\sin\chi)^2\nn \\&~&+\frac{\nu^2\sin^2\chi}{e^{2\co}}(e^{2\co}d\xi-e^{2\ao}\sin^2\theta d\Psi)^2 +\frac{e^{-8\ao-2\co}}{4\cos^2\theta}\bigg(2L^4(e^{2\ao}\nu\cos\theta\cos\chi Z_1d\theta\nn \\&~&-e^{2\co} \cos\chi V d\nu )+e^{4\ao}\cos\theta (\cos\chi d(\rho\cos\chi)-\sin\chi d(\rho\sin\chi))\bigg)^2 \bigg],\nn\\
e^{2\Phi}&\sim& \frac{256e^{6\ao-2\co}}{27\nu\rho^2(1+\frac{e^{8\ao}\cos^2\theta}{L^8})} ,\nonumber \\
B_2&\sim& \frac{3}{8}\bigg[e^{2(\ao-\co)}\sin^2\theta \cos\chi d\rho\wedge d\Psi-\rho\sin\chi d\chi\wedge d\xi\bigg]\nn \\&~& -\frac{3 L^4 e^{-8\ao-2\co}}{8\nu\rho\cos^2\theta}\bigg[e^{2\co}\rho\cos\chi V \big[ 2e^{2\co} L^4\nu \sin\chi V d\chi -e^{4\ao}\cos\theta d\rho \big]\wedge d\nu \nn \\&~&
-2 e^{4\ao+2\co}\nu^2 \cos\theta\sin\chi V (e^{2\co}\sin\chi d\rho \wedge d\xi-2e^{4\ao}\rho \cos\chi \sin^2\theta d\chi\wedge d\Psi)\nn\\&~&
+e^{2\ao}\nu \rho \cos\chi\cos\theta Z_1(e^{4\ao}\cos\theta d\rho - 2e^{2\co}L^4\nu\sin\chi V d\chi )\wedge d\theta \bigg]
\eea

Note that the first line of $B_2$ above has precisely the same structure as in the massless case.  This indicates that the NS5 intepretation is also at play here. However, as the other terms indicate, there are potentially other NS5 extended in different directions.

The RR Fluxes simplify considerably near $\mu\sim 0$:
\bea
F_1&=& \frac{9}{64L^4}\rho d\rho,\nn \\
F_3&=& \frac{9e^{-2\ao-2\co}\rho Z_1}{64}d\theta\wedge d\rho\wedge d\Psi+\frac{9e^{-6\ao}}{32L^3\cos\theta}\bigg[e^{2\co}\nu (L^8 V Z_1-2e^{8\ao}\cos^2\theta Z_2) d(\rho \cos\chi)\nn \\&~&+2e^{4\ao+2\co}\cos\theta(3f(\theta)\sin\theta-e^{4\ao}\rho\cos\theta\cos\chi Z_2)d\nu\bigg]\wedge d\theta\wedge d\Psi \nn \\&~&  +\frac{27e^{-4\ao-2\co}L^4\rho}{256\cos\theta}\bigg[e^{2\co} V (e^{2\co}\sin^2\chi d\xi+e^{2\ao}\cos^2\chi \sin^2\theta d\Psi)\wedge d\nu\nn \\&~&+\nu\sin\chi (e^{2\ao}\cos\theta\sin\chi Z_1 d\theta-Ve^{2\co}\cos\chi d\chi)\wedge (e^{2\co} d\xi-e^{2\ao}\sin^2\theta d\Psi)\bigg]\wedge d\rho \nn\\&~& + \frac{9e^{-6\ao+2\co}L^4\rho\cos\chi V Z_1}{32\cos\theta} d\theta \wedge d\nu\wedge d\Psi, \nn \\
F_5&=& \frac{27\rho}{64}d\text{Vol}(AdS_4)\wedge d\rho+\frac{9e^{-4\ao}}{32L^4\cos\theta\sin\theta f(\theta)}d\text{Vol}(AdS_4)\wedge \bigg[\nn \\&~& - e^{2\co}(3L^8 f(\theta)\sin\theta V-2e^{8\ao}\cos^2\theta Z_2)(\nu d(\rho\cos\chi)+\rho\cos\chi d\nu)\nn \\&~&+\nu^2e^{2\ao}\rho\cos\chi\cos\theta\sin\theta f(\theta)(e^{8\ao}f(\theta)\sin\theta \hat{V}+3L^8Z_1)d\theta\bigg] \\&~&
+\frac{729e^{2\ao+2\co}\rho\sin\chi}{3456 L^4}\bigg[\cos\theta\sin\chi d\nu+\nu \cos\chi\cos\theta Z_2 d\chi\bigg]\wedge d\theta\wedge d\xi\wedge d\rho\wedge d\Psi\nn.
\eea



\section{Conclusions}\label{Sec:Conclusions}

We have constructed a new solution in Type IIB with an $AdS_4$ factor and all fluxes turned on. By explicitly constructing the Killing spinor of the seed solution and further  exploiting an important result about the independence of the Killing spinor on the coordinates along which we perform NATD we were able to establish the supersymmetry of the background. Our arguments hold for both the massless case, as previously argued in \cite{Zayas:2015azn}, and now for the massive case.

Given the work on  $AdS_4$ compactifications of IIB on manifolds with local $SU(2)$ structure \cite{Lust:2009zb}, it would be extremely interesting to cast our background in this mold.
NATD acts on pure spinors, written as polyforms, via the matrix $\Omega$, which is also used to construct the dual RR fluxes.  Typically it transforms the pure spinors from SU(3) structure type to SU(2) structure type in the dual via $\slashed{\Psi}_+^{SU(2)}=i \slashed{\Psi}_-^{SU(3)}\Omega^{-1},\ \slashed{\Psi}_-^{SU(2)}=\slashed{\Psi}_+^{SU(3)}\Omega^{-1} $ \cite{Barranco:2013fza}.   We would expect to see an explicit mapping between the SO(2) family of SU(3) structures to an SO(2) family of SU(2) structures after the NATD, similar to what was shown in \cite{Lozano:2015cra}.
At the moment, simplification of the dual pure spinors into a practical form has not yielded significant results, given the shear size of the background. We, nevertheless, hope to return to this task in the future as it might shed some light on this class of backgrounds.


Finally, it would  be interesting to study the field theory dual to this class of solutions.  Much progress has been made in understanding the field theories dual to the NATD's of $AdS_5\times S^5$ \cite{Lozano:2016kum} and a reduction from M-theory of an $AdS_4$ background preserving $\mathcal{N}=4$ \cite{Lozano:2016wrs}.  
It is possible that the "completion" (as defined in those references) of the NATD backgrounds presented here fit into some $\mathcal{N}=2$ SUSY version of the $\mathcal{N}=4$ class presented in \cite{Lozano:2016wrs}.
There are, however, many new ingredients in the solutions we have constructed here.  The brane configuration setups in \cite{Lozano:2016kum,Lozano:2016wrs} hinged on the fact that the $B_2$ generated from the NATD contained a 2-cycle, on which the quantity $b_0=\frac{1}{4\pi^2}\int_{S_1^2}B_2$ is constrained to be bounded in the interval [0,1].  Identifying a similar 2-cycle in our case is not obvious, so it is at present unclear to what extent the arguments used in \cite{Lozano:2016wrs} can be generalized to our case.

\section*{Acknowledgments}
LPZ is thankful to the Abdus Salam International Centre for Theoretical Physics, Trieste, for sabbatical support during the initial stages of this project.  CAW was supported by the National Institute for Theoretical Physics of South Africa.
We would like to thank Carlos Nunez and Yolanda Lozano for useful discussions and correspondences regarding this research.

\bibliographystyle{JHEP}
\bibliography{NATD}

\end{document}